\documentclass[a4paper,10pt]{article}
\usepackage{hyperref}

\title{Imaginary Interactions with Minimum Length}
\author{ Mir Faizal$^1$ and Bhabani Prasad Mandal$^2$  \\
$^1$ Department of Physics and Astronomy, \\  University of Waterloo,   Waterloo,\\
Ontario N2L 3G1, Canada \\
$^2$ Department of Physics,  Banaras Hindu University,  \\
Varanasi-221005, India   }
 \date{}
\begin{document}
\maketitle

\begin{abstract}
We analyze the effect of having minimum length on a two dimensional anisotropic simple harmonic oscillator with PT
symmetric imaginary interaction perturbatively. First order correction to the general state is calculated analytically 
to show that it remains real as long as PT symmetry is unbroken. The characteristics of PT phase transition remain unaltered
in this deformed formulation of quantum mechanics. 
\end{abstract}

At Planck scale the picture of spacetime as a smooth manifold is expected to break down. 
This is because at Planck scale the fluctuations in the metric will be of order one. These fluctuations in the geometry  will 
give spacetime a fuzzy structure, at Planck scale,  called the spacetime foam \cite{1}-\cite{2}. 
As spacetime does not even remain a smooth manifold at Planck scale, it is difficult to study this phenomena directly. 
However, we can study the effective low energy phenomena using the fact that the Planck scale acts as a minimum length scale 
for spacetime. It may be noted that a minimum length scale also occurs naturally in   string theory \cite{z2}-\cite{2z}.
In fact, even in loop quantum gravity the big bang gets turned into a big bounce by the existence of a minimum length scale 
 \cite{z1}. There are also strong indications from black hole physics that any theory of quantum gravity should come naturally 
 equipped with a minimum length of the order of the Planck length \cite{z4}-\cite{z5}. 
Thus, a minimum length scale naturally occurs in all most all approaches to quantum gravity. However, the existence of   minimum 
length is not consistent with the usual uncertainty principle. This is because according to the usual uncertainty principle, 
one can measure the length to arbitrary accuracy if the momentum is left unmeasured. Thus, in order to make the existence of   minimum 
length consistent with quantum mechanics, the usual uncertainty principle has to be modified to Generalized Uncertainty Principle (GUP)
\cite{zasaqsw}. This modification in turn deforms the Heisenberg algebra and which leads to a different representation of the momentum 
operator in coordinate representation. 

Thus, the generalized uncertainty principle consistent with the existence of a minimum length can be written as 
$ \Delta x  \Delta p \geq \hbar \left[ 1 + \beta L_{P}^2 \hbar^{-2} \Delta p^2 \right]/2
$, where $\beta$ is a constant normally assumed to be of order unity and $L_P = 10^{-35}m$ is the Planck length. Now from this 
GUP, we obtain 
$\Delta p \leq \beta^{-1} L_P ^{-2}\hbar ( \Delta x \pm \sqrt{\Delta x^2 - \beta L_{p}^2})$. This  implies the existence of a minimum 
measurable length, $\Delta x \geq \sqrt{\beta} L_P  $. It may be noted that this algebra only becomes important 
at Planck momentum $\Delta p \sim p_L = 10^{16} TeV/c$. From now on we will adopt unites such that  $\hbar = c =1$. 
It has been argued that the minimum length can exist at an intermediate length scale between Planck length and electroweak 
length scale \cite{x1}-\cite{x2}. If this is the case, then it will be possible to detect this minimum length experimentally.  
In this connection, we will analyse the effect of minimum length on PT symmetry. 
This generalized uncertainty can be derived from a modified Heisenberg algebra   
$ [x_i, p_j ] = i [\delta_{ij} + \beta ( p^2 \delta_{ij} + 2 p_i p_j)] 
$. 
The momentum in the coordinate representation for a two dimensional system can now be represented as 
\begin{eqnarray}
 p_x &=&  - i \partial_x [1 - \beta \partial_x^2 - \beta \partial_y^2], \nonumber \\ 
  p_y &=&  - i \partial_y [1 - \beta \partial_x^2 - \beta \partial_y^2].
 \end{eqnarray}
The existence of a minimum measurable length deforms 
all quantum mechanical Hamiltonians  by 
$
 H \psi = H_0\psi + H_1\psi
$, where $H_0 =    -(\partial^2_x + \partial^2_y) /2m + V (x, y) $ is the original Hamiltonian, 
and 
$H_1 = \beta    (\partial_x^4 + \partial_y^4  + 2 \partial_x^2 \partial_y^2) /m$ 
is the term that occurs due to the existence of a  minimum length. 
This deformation can leads to the existence of non-Hermitian operators \cite{a1111}.
Thus, when studying the systems with minimum length, there is no reason to restrict the interactions to 
have real coefficients. This is because the non-Hermitian behavior can also be introduced even in the systems 
by the existence of a minimum length. In this paper we will study a 2D anisotropic simple harmonic oscillator with PT symmetric 
imaginary interaction in this deformed quantum theory.

 We start with the following Hamiltonian of a 2D anisotropic SHO\cite{b}, 
\begin{equation}
 H =  \frac{1}{2}[p_x^2 +  p_y^2  +  m \omega_x^2 x^2 +  m \omega_y^2 y^2  +2 i \lambda xy],
\label{ham}
\end{equation}
where $\lambda$ is real and $\omega_x \neq \omega_y$. 
The imaginary interaction  $i\lambda xy$ is PT symmetric as in 2D, parity is defined as either
of $ x\rightarrow -x ; \  y\rightarrow y $ or $ x\rightarrow x ; \  y\rightarrow -y $. The energy eigenvalues and eigenfunctions for this system are obtained by solving the Schroedinger equation corresponding to this system in usual quantum mechanics as 

\begin{equation}
 E_{n_1, n_2} = \left(n_1 + \frac{1}{2}\right)   c_1 + \left(n_2 + \frac{1}{2}\right)   c_2,
 \label{eng}
\end{equation}
and 
\begin{equation}
 \psi_{n_1, n_2} (X, Y) = N \exp \left( - \frac{\alpha_1 X^2}{2} - \frac{\alpha_2 Y^2}{2}\right) H_{n_1} (\alpha_1 X) H_{n_2} (\alpha_2 Y), 
\end{equation}
where 
\begin{equation}
 N =\sqrt{ \frac{\alpha_1\alpha_2}{ n_1 ! n_2 ! \pi  2^{n_1 + n_2}}},
\end{equation}
 is a normalization constant. Here $c_1$ and $c_2$ are given by 
\begin{eqnarray}
 c_1^2 = \frac{1}{2}\left[ \omega_+ ^2 -\frac{\omega_-^2}{k}\right], \nonumber \\
 c_2^2 = \frac{1}{2}\left[ \omega_+ ^2 +\frac{\omega_-^2}{k}\right], 
\end{eqnarray}
where 
$ \omega_\pm^2 = \omega_y^2 \pm \omega_x^2$, $k^{-1} = \sqrt{1 -  {4 \lambda^2/m^2 \omega_-^4}}$,
and  $\alpha_1^2 = m c_1, $ and $ \alpha_2^2 = m c_2$.
The coordinates in a new basis are defined as 
\begin{eqnarray}
  X &=& \sqrt{\frac{1 + k}{2}}x -  \sqrt{\frac{1 - k}{2}}y,  \nonumber \\ 
  Y &=& \sqrt{\frac{1 - k}{2}}x -  \sqrt{\frac{1 + k}{2}}y. 
\end{eqnarray}

Now when  $\kappa $ is real i.e. $|\lambda|\le |\frac{m\omega_-^2}{2}|\equiv \lambda_c $ the parameters
$c_1$ and $c_2$ are real and the entire spectrum in Eq. (\ref{eng}) is real
and the system is in unbroken PT phase as in this situation 
\begin{eqnarray}
PT\psi_{n_1,n_2}(x,y) = (-1)^n_1 \psi_{n_1,n_2}(x,y),\nonumber\, \, \, \,  \mbox {or,  }\\
PT\psi_{n_1,n_2}(x,y) = (-1)^n_2 \psi_{n_1,n_2}(x,y).
\label{2pt}
\end{eqnarray}

However when $|\lambda|> |\frac{m\omega_-^2}{2}|$, \ $\kappa $ is imaginary and we have
complex spectrum occurring in conjugate pairs. The system is in broken PT phase as the wave function $\psi_{n_1,n_2}(x,y)$ in this situation does not  respect the PT symmetry.
The critical value of the the coupling, $\lambda_c$ depends on the anisotropy, i.e. $\omega_-$ of the system. If the system is more anisotropic the span of the PT unbroken phase is longer and the system remains in broken PT phase all the time when the system becomes isotropic, i.e. $ \omega_-=0 $ and hence$\lambda_c=0$

Now deforming this Hamiltonian using GUP, we obtain  
\begin{equation}
 H = H_0 + H_I, 
\end{equation}
where $H_0$ is the original Hamiltonian as written in Eq. (\ref{ham}) with PT symmetric imaginary interaction and the effect of deformation due to GUP is
\begin{equation}
 H_I = \frac{\beta}{m} [\partial_x^4 + \partial_y^4  + 2 \partial_x^2 \partial_y^2].  
\end{equation}
We treat this new interaction term perturbatively to calculate
the corrections to $E_{n_1, n_2}$ and $\psi_{n_1, n_2}(x,y)$ .  First we observe that 
$\partial_x^4 + \partial_y^4  + 2 \partial_x^2 \partial_y^2 = \partial_X^4 + \partial_Y^4  + 2 \partial_X^2 \partial_Y^2$.
Now using the relation
\begin{eqnarray}
 \frac{d^4}{d s^4} \exp(-s^2/2) H_n (s) &=& [ (s^4 - 6  s^2 +3 ) H_n(s)  - 4 (s^3 - 3 s) H'_n (s) \nonumber \\ &&  + 
 6(s^2 -1) H_n '' (s) - 4 s H'''_n (s)  + H_n ''''(s)  ] \nonumber \\ && \times \exp (-s^2/2),
\end{eqnarray} and properties of Hermite polynomial, $H_n(x)$
we obtain the following result 

\begin{eqnarray}
 \Delta E_{n_1, n_2} &=& <\psi_{n_1,n_2}|H_I|\psi_{n_1,n_2}>\nonumber \\ 
 &=& \frac{\beta}{2m} \Big[ 3 \left(n_1^2 + n_1 + \frac{1}{2}\right) \alpha_1^4 + 3 \left(n_2^2 + n_2 + \frac{1}{2} \right)\alpha_1^4 
 \nonumber \\  &&  + 
 (2 n_1 + 1)(2 n_2 + 1) \alpha^2_1 \alpha^2 _2  \Big].  
 \label{eqns}
\end{eqnarray}and some of the special cases are
\begin{eqnarray}
 \Delta E_{0,0} &=& \frac{\beta}{2m} \left[\frac{3}{2} \alpha_1^4 + \frac{3}{2} \alpha_2^4   + \alpha_1^2 \alpha^2_2 \right], \nonumber \\ 
  \Delta E_{0,1} &=& \frac{\beta}{2m} \left[\frac{3}{2} \alpha_1^4 + \frac{15}{2} \alpha_2^4   + 3\alpha_1^2 \alpha^2_2 \right], \nonumber \\ 
 \Delta E_{1,0} &=& \frac{\beta}{2m} \left[\frac{15}{2} \alpha_1^4 + \frac{3}{2} \alpha_2^4   + 3\alpha_1^2 \alpha^2_2 \right], \nonumber \\ 
  \Delta E_{1,1} &=& \frac{\beta}{2m} \left[\frac{15}{2} \alpha_1^4 + \frac{15}{2} \alpha_2^4   + 9\alpha_1^2 \alpha^2_2 \right].
\end{eqnarray}
The first order corrections to  general eigenfunctions can easily be calculated using 
\begin{equation}
\Delta\psi_{n_1,n_2}(x,y)= M_{m_1, m_2, n_1, n_2 }\times \psi_{m_1,m_2}(x,y), 
\label{wav}
\end{equation}
where 
\begin{equation}
  M_{m_1, m_2, n_1, n_2 } = \sum_{\{m_1,m_2\}\neq \{n_1, n_2)\}}
  \frac{<\psi_{m_1,m_2}(x,y)|H_I|\psi_{n_1,n_2}(x,y)> }{E_{n_1,n_2}-E_{m_1,m_2}}. 
\end{equation}

Now we have the following interesting observations 

(i) When $|\lambda|\le |\lambda_c |$ the parameter $\kappa $ is real hence $\alpha_1, \alpha_2$ are also real. The first order correction to the energy eigenvalues, $\Delta E_{n_1,n_2}$
due to minimum length deformation is real. Now we argue that the deformed system  remains in the PT unbroken phase in this condition. To show that we have to make sure $\Delta\psi_{n_1,n_2}(x,y)$  will have same PT phase with $\psi_{n_1,n_2}(x,y)$. Now it is easy to verify that right hand side of Eq. (\ref{wav}) changes under PT as $(-1)^{n_1}$ or  $(-1)^{n_2}$ which is exactly same as the PT properties of $\psi_{n_1,n_2} (x,y)$ as given in Eq. (\ref{2pt}).

(ii)However for  $|\lambda|> |\lambda_c| $ the corrections due to minimum length deformation becomes complex and the deformed theory moves to PT broken phase as the first order correction to wave function,  $\Delta\psi_{n_1,n_2}(x,y)$ does not respect PT symmetry.

(iii) At the critical coupling,  $|\lambda|= |\lambda_c| $,    
 $\kappa ^{-1}=0$ , and $\alpha_1= \alpha_2 $, additional degeneracy occurs as $E_{n_1,n_2} = E_{n_2, n_1}$. This fact is also reflected in the deformed theory.

(iv) When the system is isotropic $\omega_x = \omega_y $ i.e. $\lambda c =0 $, then the restriction $|\lambda| \leq  |\lambda_c|$ never holds and the system is 
always in the broken phase. So, in this case the corrections are always complex as we see from the Eq. (\ref{eqns})

(v)Finally,  when $n_1 = n_2$, the energy correction $\Delta E_{n_1, n_2}$ is real and 
the PT symmetry is unbroken even for $|\lambda|>|\lambda_c|.$

 We have  analyzed the effect of deforming a two dimensional anisotropic harmonic oscillator with PT symmetric imaginary coupling  by 
GUP. The characteristics of PT phase transition remain unaltered in this deformed formulation perturbatively. 
It would be interesting to analyze other non-Hermitian systems further using GUP. 
In fact, GUP with a linear term in momentum has also been  introduced  \cite{y1}. 
This leads to non-locality in any  dimension beyond the simple  one dimensional case \cite{y2}.
However, for one dimensional systems there is no such non-locality. 
It would be interesting to analyze one dimensional Hermitian systems using this term. 

\vspace{.2in}

{\bf Acknowledgments:} One of us (BPM) acknowledges the financial support from a Department of Science and Technology (DST), Government of India, under the SERC project sanction grant No: SR/S2/HEP-0009/2012.


\begin{thebibliography}{99}
\bibitem{1}  S. W. Hawking, Phys. Rev. D 37,  904 (1988) 
\bibitem{Hawking88} S. W. Hawking and D. N. Page, Phys. Rev. D 42,  2655  (1990)
\bibitem{1111}M. Faizal, JETP. 114, 400 (2012)
\bibitem{colm}S. Coleman, Nucl. Phys. B 310, 643 (1988)
\bibitem{2}S. Cokman, Nucl. Phys. B 307,  867 (1988)
\bibitem{z2}D. Amati, M. Ciafaloni and G. Veneziano, Phys. Lett. B 216, 41 (1989)
\bibitem{csdcas}L.  N.  Chang, D.  Minic, N. Okamura, and T.  Takeuchi, Phys.Rev. D65,  125027 (2002)
\bibitem{cscds}L.  N.  Chang, D.  Minic, N. Okamura, and T.  Takeuchi, Phys. Rev. D65, 125028 (2002) 
\bibitem{2z}S. Benczik, L. N.  Chang, D.  Minic, N.  Okamura, S.  Rayyan, and T.  Takeuchi,  Phys. Rev. D66, 026003 (2002)
\bibitem{z1}P.  Dzierzak, J.  Jezierski, P.  Malkiewicz, and W. Piechocki,  Acta Phys. Polon. B41, 717 (2010) 
\bibitem{z4} M. Maggiore, Phys. Lett. B304, 65 (1993)
\bibitem{z5}M. I. Park, Phys. Lett. B659, 698 (2008) 
\bibitem{x1}A. F.  Ali, S. Das and E.  C. Vagenas,  Phys. Rev. D 84, 044013 (2011)
\bibitem{x2}   S. Das and E. C. Vagenas,  Phys. Rev. Lett. 104, 119002 (2010) 
\bibitem{zasaqsw}  A. Kempf, G. Mangano, and R. B. Mann, Phys. Rev. D 52, 1108 (1995)
\bibitem{a1111} B. Bagchi and A. Fring,  Phys. Lett. A 373, 4307 (2009)  
\bibitem{b} B. P.  Mandal , B. K.  Mourya and  R. K.  Yadav,  Phys.  Lett.  A 377, 1043 (2013) 
\bibitem{y1} A. F. Ali, S. Das and E. C. Vagenas, Phys. Lett. B 678, 497 (2009) 
\bibitem{y2}M. Faizal,   Int. J. Geom. Meth. Mod. Phys. 12, 1550022 (2015)
\end{thebibliography}
\end{document}